\documentclass[aps,prd,10pt,onecolumn,nofootinbib,showpacs,showkeys,notitlepage]{revtex4-1}
\usepackage{amsmath,graphicx,bm}
\usepackage{hyperref}

\newcommand{\cH}{{\cal H}}

\renewcommand{\d}{\partial}

\newcommand{\rh}{\varrho}

\newcommand{\ep}{\varepsilon}

\newcommand{\p}{{\bf p}}

\renewcommand{\Im}{\,\textrm{Im}\,}

\newcommand{\pint}[2]{{\int\!\frac{d^{#1}#2}{(2\pi)^#1}\,}}
\newcommand{\pintz}[1]{{\int\!\frac{d #1}{2\pi}\,}}

\newlength{\szovszel}\newlength{\szovmag}

\newcommand\lsim{\mathrel{\rlap{\lower4pt\hbox{\hskip1pt$\sim$}} \raise1pt\hbox{$<$}}}                
\newcommand\gsim{\mathrel{\rlap{\lower4pt\hbox{\hskip1pt$\sim$}} \raise1pt\hbox{$>$}}}                

\newcommand{\K}{{\cal K}}

\begin{document}

\title{Hadron melting and QCD thermodynamics}

\author{A. Jakov\'ac} 
\affiliation{Institute of Physics, Eotvos University, H-1117 Budapest, Hungary}

\date{\today}

\begin{abstract}
  We study in this paper mechanisms of hadron melting based on the
  spectral representation of hadronic quantum channels, and examine
  the hadron width dependence of the pressure. The findings are
  applied to a statistical hadron model of QCD thermodynamics, where
  hadron masses are distributed by the Hagedorn model and a uniform
  mechanism for producing hadron widths is assumed. According to this
  model the hadron - quark gluon plasma transition occurs at $T\approx
  200$-$250$ MeV, the numerically observable $T_c=156$ MeV crossover
  temperature is relevant for the onset of the hadron melting process.
\end{abstract}

\maketitle

\section{Introduction}
\label{sec:intro}

Hadron Resonance Gas\cite{HRG0} (HRG) provides a good description for
a large number of low-temperature QCD observables, as it was
demonstrated by a number of studies comparing the result of Monte
Carlo (MC) lattice simulations of QCD and the HRG predictions
\cite{Andronic:2003,Karschetal,Huovinen:2009yb,Borsanyi:2010cj,Bazavov:2013dta}.
The strategy of the HRG is quite simple: one takes the particle masses
reported by the Particle Data Group \cite{PDG} and plug them into the
ideal gas expression of the pressure. In case of simulations at
non-physical masses one has to adjust the hadron masses to the actual
numerical environment \cite{Huovinen:2009yb}.

The success of the HRG approach, however, raises fundamental questions
about the QCD phase transition. The traditional physical reasoning is
based on the \emph{entropy} argument. It states that in the quark
gluon plasma (QGP), because of the color of the quarks, the number of
degrees of freedom is much larger than that in the hadronic sector, if we
count only the stable hadrons (in a 2-flavor case we have 3 pions and
2 nucleons, 3 bosonic and 8 fermionic degrees of freedom, while the
2-flavor QGP contains 8 gluons and $2\times 3$ quarks, i.e. 16 bosonic
and 24 fermionic degrees of freedom). Then the QGP phase, although
energetically less favorable, has larger entropy, and thus at high
temperature, where the entropy dominates the free energy, this will be
the thermodynamic ground state of QCD.

This appealing picture is questioned by the success of HRG, in which
\emph{all hadrons} are treated as thermodynamical degrees of freedom
on equal footing. To be more precise, to describe the pressure
measured by Monte Carlo (MC) simulations \cite{Borsanyi:2010cj} up to
the reported crossover temperature $T_c\approx156$ MeV
\cite{Aoki:2006we}, one has to take into account ${\cal O}(1000)$
hadronic resonances, meaning that in the hadronic phase the effective
number of degrees of freedom at this temperature is of this order. But
if it is so, the hadron phase is not just energetically favorable, but
also has larger entropy, and then no phase transition would be
allowed.

It would help if, because of deconfinement, the hadron phase became
unstable at a certain temperature, as it was suggested by Hagedorn
\cite{Hagedorn:1965st}. But, according to MC results, the transition
at $T_c=156$ MeV is a smooth crossover, and so there is no room for
such a violent process like the collapse of a phase. Another hint that
the hadron phase will not be unstable above $T_c$ is that one can
identify the hadronic states at $T>T_c$, even up to $T\sim
1.5$-$2\,T_c$ \cite{Datta:2003ww,Umeda:2002vr,Asakawa:2003re,
  Jakovac:2006sf, Petreczky:2012ct}! In a recent MC study
\cite{Bellwied:2013cta} the authors state that the crossover
temperature is significantly different for strange and non-strange
hadrons, thus there can not be any specific temperature where the
hadron phase collapses. All this suggests that the hadrons are present
above $T_c$ -- with their huge entropy factor which forbids a phase
transition! Therefore the question is that \emph{what happens at
  $T_c$}?

To be able to answer this question one has to study the hadrons, since
they give the dominant part of the entropy at this temperature. There
are several facts warning us that the original HRG description is not
enough at these temperatures \cite{Huovinen:2009yb}. The most
pertinent property of the naive HRG is that after a certain
temperature (about $200$ MeV) it starts to seriously overestimate the
real pressure. Strangeness correlations also cannot be described
correctly above $T=150$-$170$ MeV (depending on the correlator) with
the uncorrelated HRG model \cite{Bazavov:2013dta,
  Bellwied:2013cta}. Moreover, a weekly interacting gas would yield
large (dimensionless) transport coefficients, in contrast to the
reported small value of $\eta/s$ \cite{Heinz:2013th}.

It is clear that HRG is the simplest representation of the hadron
spectrum, in reality it expresses much richer structures than just
stable hadrons. Most prominently the hadrons are quasiparticles, and
have a finite width, typically of the order ${\cal O}(100)$
MeV. Moreover the hadron spectrum contains the continuum of scattering
states of other hadrons and eventually of quarks. These properties can
be important, in principle the complete spectrum may influence the
thermodynamics. For example we expect that hadron width increases with
temperature, and finally the hadrons will disappear in the continuum
of scattering states: the hadron \emph{melts}.

There exist different approaches in the literature to take into
account the effect of the hadronic width for thermodynamics
\cite{Blaschke:2003ut, Biro:2006zy, Biro:2006sv} and also in transport
phenomena \cite{Ivanov:1998nv,Ivanov:1999tj,Peshier:2005pp}. These
approaches use the approximation that all hadronic resonances are
independent. This assumption, although correct for well separated
particle peaks \cite{RD}, gets into trouble when the quasiparticle
peaks in a given quantum channel\footnote{Quantum channel in this work
  means superselection class, i.e. states having the same quantum
  numbers.}  start to overlap, or when a peak is too close to a
multi-particle threshold. In the S-matrix approach this is discussed
in Refs.~\cite{Scon,Svec}. To understand the problem qualitatively, we
recall that a quasiparticle is a collective mode, i.e. it is a mixture
of the original one-particle energy level and various (actually
infinitely many) multiparticle configurations. If two peaks overlap,
both quasiparticles would contain the states in the overlapping
region. Then, if we performed calculations with two independent
quasiparticles, we would overcount the states in the common region. We
encounter a similar situation when a peak is too close to a continuum:
the overlap would be counted twice if we assumed that the
quasiparticle represents a full thermodynamical degree of
freedom. This qualitative picture suggest that in these cases we have
to decrease the effective number of degrees of freedom.

The goal of this paper is to give a systematic description of this
process, referred to as hadron melting, find its effect on the
thermodynamics and finally to give some answer to the above question
that ``what is going on at $T_c$''. Our investigation is based on a
previous study, Ref.~\cite{Jakovac:2012tn} where the general,
mathematically consistent strategy of treating excitations with
arbitrary spectral function was worked out, without separating them in
an artificial way into quasiparticles, for determining thermodynamical
quantities. Here we try to apply these ideas to QCD.

A big obstacle in front of an accurate description of the hadron gas
is that the spectral details are usually not known for the hadronic
channels, in particular not in the case of large mass hadrons. But, in
fact, for thermodynamics we just need a statistical description of the
hadrons. Therefore, in this paper, we use an idealized description,
based on the Hagedorn spectrum \cite{Hagedorn:1965st}. In this
approach one estimates the density of hadronic states by an
exponentially growing function. The original proposal of Hagedorn was
$\rh_{hadr}(m) \sim (m^2+m_0^2)^{-5/4} e^{m/T_H}$, where $m_0\approx
500$ MeV and $T_H$, the Hagedorn temperature, is of the order of
$200-300$ MeV. The validity of this assumption was verified lately
\cite{Broniowski:2004yh}, where the authors have pointed out that a
large variety of exponentially growing functions (in particular a pure
exponential) is appropriate to fit the observed hadron
spectrum. Similar works prove the plausibility of the
Hagedorn-assumption \cite{NoronhaHostler:2012ug}. In an ensemble
containing free particles with masses distributed by the Hagedorn
distribution, the pressure has a singularity at $T=T_H$. This
resembles to the fast growth of the pressure in the HRG computations:
this is also due to the contribution of the large numbers of massive
hadronic states. Therefore, to understand the reduction of the
pressure of the real hadron gas, the study of the Hagedorn gas is an
excellent tool.

Our strategy is to first study simplified models of melting of a bound
state, focusing on the possible reasons of melting: a particle can
melt into a continuum because of growing width, and because of
decreasing wave function renormalization constant. In principle there
can be a third mechanism, when different quasiparticle peaks merge to
form a broad continuum. This can be relevant for the Coulomb problem
of QED, for example, where infinitely large number of states are
present very densely at zero temperature in each quantum channel
(eg. in the $s$ channel the $ns$ states have energy $E_0/n^2$ and
width $\sim1/n^3$). But, according to the PDG tables, in QCD in each
quantum channel there are at most 2-3 excitation peaks present, so
probably the most important source of melting of hadrons is one of the
two scenarios mentioned above. The lesson of the model studies will be
summarized in a single simple formula, giving the dependence of the
pressure of the system on the quasiparticle width, in the presence of
a continuum. This is a key point of the paper,
cf. \eqref{eq:meltansatz}. The coefficient for the reduction of the
effective number of degrees of freedom reads $\sim e^{-c\gamma^2(T)}$,
where $\gamma(T)$ is the hadron width. This formula makes it possible
to describe, how in the course of hadron melting the pressure of the
hadronic matter decreases. We find that up to $200-250$ MeV one can
easily describe the existing MC data even with the simplest purely
hadronic fits, but above this temperature, in all parametrizations we
used, the hadronic pressure contribution becomes much smaller than
the measured one.

This leads us to propose the following answer to the above posed
question: \emph{$T_c=156$ MeV is the starting point of rapid
  hadron melting, and not a hadron - QGP transition}. The quark gluon
plasma appears at a higher $T'_c$ temperature; our model calculations
propose for it a range of $T'_c\in[200,250]$ MeV.

The paper is organized as follows: first we shortly overview the
consequences of the Hagedorn spectrum for the energy density and
pressure, and recall the most important results of
Ref.~\cite{Jakovac:2012tn}. Next we will consider the two simplified
models of particle melting. We study the energy density and pressure
in these models, and in particular we determine the effective number
of degrees of freedom, i.e. the pressure reduction factor. Then we
apply the findings to QCD thermodynamics, and present our fit to the
results of the MC measurements. We close the paper with our Conclusions.

\section{Particle distribution and spectral functions}
\label{sec:part}

If we have free bosonic/fermionic particles with masses $m_n$
($n=1\dots N$), then the pressure of this mixture (at zero chemical
potentials) reads in thermal equilibrium
\begin{equation}
  \label{eq:Hformula}
  P_\alpha = -\alpha\frac T{2\pi^2} \sum\limits_{n=1}^N \int\limits_0^\infty
  dp\, p^2\,\ln\left(1-\alpha e^{-E(p,m_n)/T}\right),
\end{equation}
where $\alpha=\pm1$ for bosons/fermions, and $E(p,m)$ is the
dispersion relation, for relativistic case $E^2=p^2+m^2$.

For a large number of nearby mass values we can introduce the
(hadronic) mass density function $\rh_{hadr}$ with the definition
\begin{equation}
  \rh_{hadr}(m_n) = \frac1{m_{n+1}-m_n}.
\end{equation}
This makes possible to approximate the above formula as an integral,
formally inserting $dm \rh(m)=1$ and apply Riemann integral formula
\begin{equation}
  P_\alpha \approx -\alpha\frac T{2\pi^2}
  \int\limits_0^\infty\!dm\int\limits_0^\infty dp\, \rh_{hadr}(m)
  p^2\,\ln\left(1-\alpha e^{-E(p,m)/T}\right).
\end{equation}
For massive hadrons $m\gg T$ we can use nonrelativistic approximation
$E=m+p^2/(2m)$, and $e^{-E/T}\ll1$. Then we find
\begin{equation}
  P\approx \frac{T^{5/2}}{(2\pi)^{3/2}} \int\limits_0^\infty\!dm\,\rh_{hadr}(m)
  m^{\frac32} e^{-m/T},
\end{equation}
In the case of Hagedorn spectrum the hadronic density function grows
exponentially $\rh_h(m)\sim e^{m/T_H}$, and the integral is divergent
for $T>T_H$. This would signal, according to the original
argumentation of \cite{Hagedorn:1965st}, a thermodynamical
instability of the hadron gas. This scenario, however, is manifested
in QCD in a different way.

\begin{figure}[htbp]
  \centering
  \includegraphics[height=5cm]{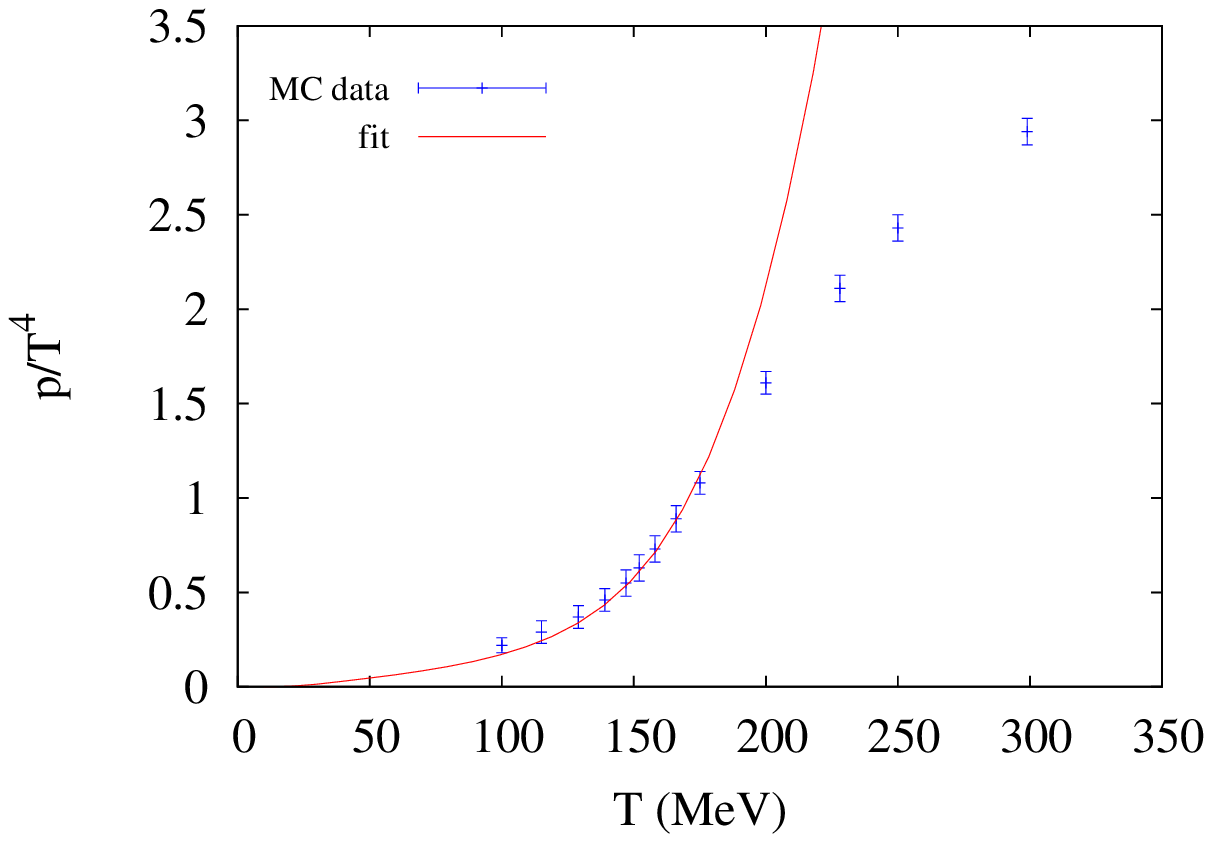}
  \includegraphics[height=5cm]{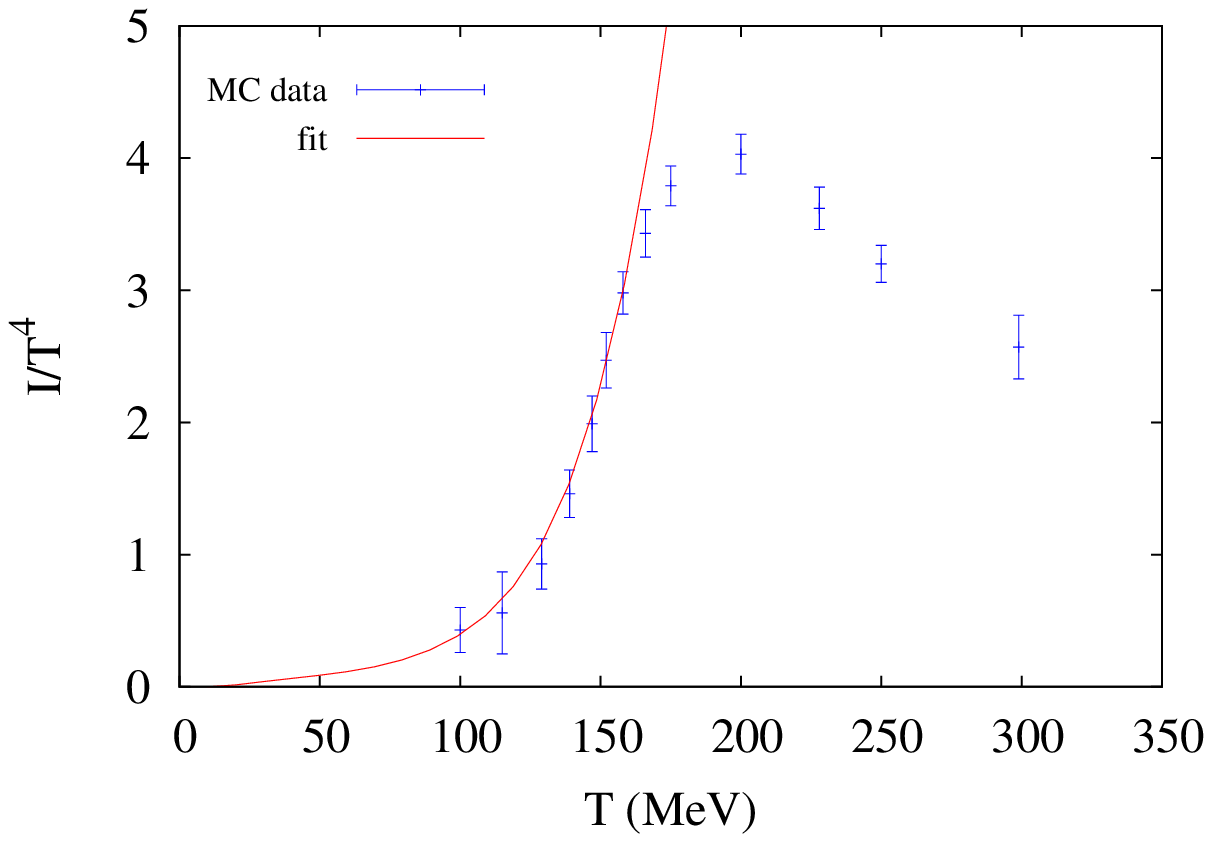}
  \hspace*{3em}a.)\hspace*{5cm}b.)
  \caption{Hagedorn fits to the MC pressure data (a.) and interaction
    measure ($I=\ep-3p$) data (b.) from \cite{Borsanyi:2010cj}. The
    solid (red) line is the result of 5000 particles with Hagedorn
    mass distribution \eqref{eq:masslevels} with parameters $M=120$
    MeV and $T_H=240$ MeV.}
  \label{fig:MCfit}
\end{figure}

For numerical computation we should use the discrete mass summation of
eq.~\eqref{eq:Hformula}. For simplicity we apply a purely exponential
hadron mass density $\rh_{hadr}$ which corresponds to the following
hadron mass series
\begin{equation}
  \label{eq:masslevels}
  m_n = M +T_H\ln n.
\end{equation}
The parameters ($M$ and $T_H$) of this formula can be chosen to fit
the pressure data from Monte Carlo (MC) simulations. In
Fig. \ref{fig:MCfit} we used 5000 bosonic resonances to fit the data
of the BMW group \cite{Borsanyi:2010cj} for the pressure and the
interaction measure or trace anomaly ($I = \ep-3p$), both scaled
by $T^4$.
The fit parameters turned out to be $M=120$ MeV and $T_H=240$
MeV. Both are reasonable values, $M$ is close to the pion
mass\footnote{In Hagedorn spectrum all hadrons participate without
  multiplicity factor, so real pions should be distributed at three
  different mass values.} and a reasonable Hagedorn temperature
value. It is remarkable that even with this simplest mass density
formula we obtain excellent fits at lower temperatures. Since the
number of excitations is finite, we do not have a real singularity at
$T_H$, though the value of the corresponding SB limit is still about
50 times the SB value of QCD.

In the above analysis we have only taken into account the free gas
description of the hadronic matter. The high mass resonances, however,
usually have very large width, often in the range of ${\cal O}(100)$
MeV which is temperature dependent, and they are close to the
continuum formed by hadronic scattering states and -- at higher
temperatures -- by quark-gluon scattering states. So, in particular
at growing temperature, sooner or later we inevitably get into the
regime when the resonances in a given quantum channel overlap with
each other and with the multiparticle continuum. Then, as it was
discussed in the introduction, the free gas approximation
significantly overestimates the real pressure, since the overlapping
regimes are counted more than once.

The correct treatment of a system described by a generic spectral
function $\rh_S(p)$ was worked out in \cite{Jakovac:2012tn}. The key
result of that paper is that the energy density and pressure coming
from a bosonic excitation should be calculated by the formulae
\begin{equation}
  \label{eq:master}
  \ep = \pint4p \rh_S(p) \cH(p) \Theta(p_0) n(p_0),\qquad 
  P = -T \pint4p \rh_S(p) \cH(p) \frac{\Theta(p_0)}{p_0}
  \ln\!\left(1-e^{-\beta p_0}\right),
\end{equation}
where $n(p_0)=(e^{\beta p_0}-1)^{-1}$ is the Bose-Einstein
distribution, and (with a principal value integration denoted by
${\cal P}$):
\begin{equation}
  \K^{-1}(p) = {\cal P}\pintz\omega \frac{\rh_S(\omega,\p)}{p_0-\omega},\qquad
  \cH(p) = p_0 \frac{\d\K}{dp_0} -\K.
\end{equation}
An important property of \eqref{eq:master} is that the thermodynamical
quantities are insensitive to the normalization of the spectral
function, since $\K$ and thus $\cH$ is inversely proportional to
$\rh_S$. Therefore, although the proper magnitude of $\rh_S$ is
determined by the sum rules, its actual value does not influence
thermodynamics.

Technically one arrives at these formulae by introducing an effective
quadratic theory with kernel $\K$ which reproduces the input spectral
function $\rh_S(p)$. Then one calculates the energy-momentum tensor
(Noether-current), and takes the finite temperature expectation
values. The equations are quite plausible, for example the energy
density $\ep$ sums up, for all spatial momenta, all excitation energy
levels with a Bose-Einstein distribution, associating $\cH$ energy
density contribution to them. The distinction relative to the naive
formula is that here the associated energy density also depends on the
spectral function in a nontrivial way. Heuristically, $\cH$ is related
to the kernel of the quadratic effective theory $\K$, like the
Hamiltonian is related to a Lagrangian in an ordinary mechanical
system, if $p_0$ replaces the time derivative of the coordinate.

It was shown in \cite{Jakovac:2012tn} that if $\rh_S$ consists of
Dirac-delta peaks
\begin{equation}
  \label{freespectrum}
  \rh_S(p)\biggr|_{free} = \sum_{n=1}^\infty Z_n \delta(p^2-m_n^2),
\end{equation}
then eq.~\eqref{eq:master} reproduces the usual ideal gas energy
density of independent free particles. It is remarkable that in this
limit the thermodynamics is independent of the individual
normalizations $Z_i$. This, of course, will not be true any more in
case of overlapping peaks, it is only invariant under an overall
normalization of the spectral function.

\section{Models of hadron melting}
\label{sec:melt}

Now the task is to go beyond the ideal gas description of
\eqref{freespectrum}, and try to incorporate the width and the
threshold effects into the spectral function. For a realistic
description we would need a detailed knowledge of the hadronic
spectral functions at finite temperature. This is not available yet,
and probably it will be not available in the near future. Therefore
here we consider two clean model-scenarios, corresponding to different
mechanisms of hadron melting. In both cases we will study the
thermodynamics of the system.

In these scenarios we consider a spectral function describing a
quasiparticle peak at mass $m$ and a multiparticle continuum. The
nonzero value of the continuum at the quasiparticle energy yields a
finite width for the quasiparticle. The most general setup would have
been a model consisting of a finite number of bound states with
different width parameters, and a multiparticle continuum. According
to the PDG particle tables, however, in most quantum channels there is
only one observed bound state, in particular in the higher mass
regions, but also in case of smaller masses there are just a few
(2-3). Therefore, the simplest and quite realistic approach is to
study the thermodynamics when a single bound state peak overlaps with
the continuum states.

In one of the scenarios we will vary the width of the
quasiparticle. At zero temperature it can be even zero if the
threshold of the continuum $m_{th}$ is larger than the particle mass
$m$. This describes a stable particle, it then represents one
thermodynamical degree of freedom. As the temperature increases, a
nonzero value of the spectral function develops below the zero
temperature threshold which yields nonzero width for the bound
state. Finally, when the width is large, the bound state peak melts
into the continuum. Physically this situation can be the most relevant
when studying bound states near the threshold. Note, that when the
width is large enough, the bound state is not identifiable at all, it
completely dissociates into its components forming the continuum.

In the other scenario we consider a bound state which may have a small
width, but the main process is the decrease of the wave function
renormalization of the peak. This can happen because the multiparticle
continuum increases by temperature effects. Because of the sum rule
this must be accompanied by a shrinking quasiparticle wave function
renormalization. As a result, the particle sinks into the
continuum, and disappears from the thermal medium, even when its width
is not changing too much. In this case the quasiparticle remains
identifiable, just its thermal weight gets smaller. This situation is
manifested in usual chemical reactions, when all reagents keep their
(quasi)particle nature, just their concentration changes. 

In the melting of hadrons probably both effects are present. In MC
simulations \cite{Jakovac:2006sf} one can clearly see the decreasing
height of bound state peaks (e.g. in $J/\Psi$ channel), but the
spectral representation is not precise enough to make distinction
between the two possible scenarios.

For both scenarios, the quasiparticle peak is represented by a
Lorentzian:
\begin{equation}
  \label{eq:Lorformula}
  \rh_{Lor}(p) = \frac{4 h\gamma^2 p^2}{(p^2-m^2)^2 + 4\gamma^2 p^2}.
\end{equation}
In this parametrization $h$ is the height of the peak. Near $p^2=m^2$
mass shell the above formula behaves as
\begin{equation}
  \rh_{Lor}(p=m+x) \approx \frac{\zeta\gamma}{x^2+\gamma^2},\qquad
  \zeta=h\gamma,
\end{equation}
and so $\gamma$ is the peak width, $\zeta$ is the wave function
renormalization constant.

The continuum is modeled by a 2-particle spectral function, but with
imaginary threshold mass (in order to mimic its value under the
threshold value which is responsible for the width of the
quasiparticle peak), and a correction prefactor to improve the $p\to0$
behavior:
\begin{equation}
  \rh_{cont}(p) = \frac p{p^2+m_{th}^2} \Im\sqrt{m_{th}^2+iS-p^2} =
  \frac p{p^2+m_{th}^2} \sqrt{\frac{\sqrt{(p^2-m_{th}^2)^2+S^2} +
      p^2-m_{th}^2}2 }.
\end{equation}
Its normalization is chosen to be unity, therefore the quasiparticle
$\zeta$ is a relative wave function renormalization. The complete
spectral function is $\rh=Z(\rh_{Lor}+\rh_{cont})$, where the value of
$Z$ is responsible for satisfying the sum rule. Then the complete wave
function renormalization of the quasiparticle peak is $\zeta Z$. But
$Z$, being an overall normalization factor, drops out from
thermodynamical observables, as it was discussed above.

In the present paper we used $m=1$ and $m_{th}=2$ values. $S$ is
determined by the condition that the value of the continuum at $m$ is
$\gamma$.  In the first scenario we keep $\zeta$ fixed and increase
the peak width, in the second scenario we keep $\gamma$ fixed, and
decrease the peak height.

\begin{figure}[htbp]
  \centering
  \includegraphics[height=5cm]{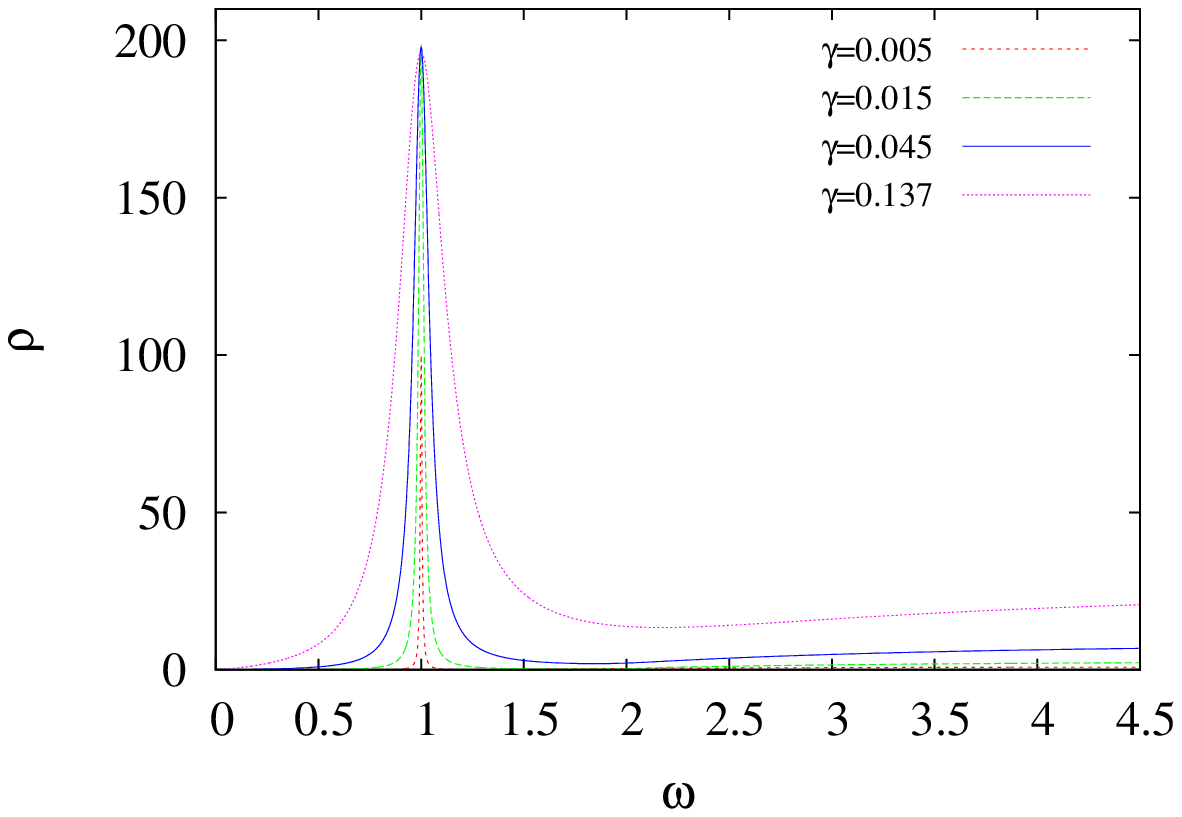}
  \includegraphics[height=5cm]{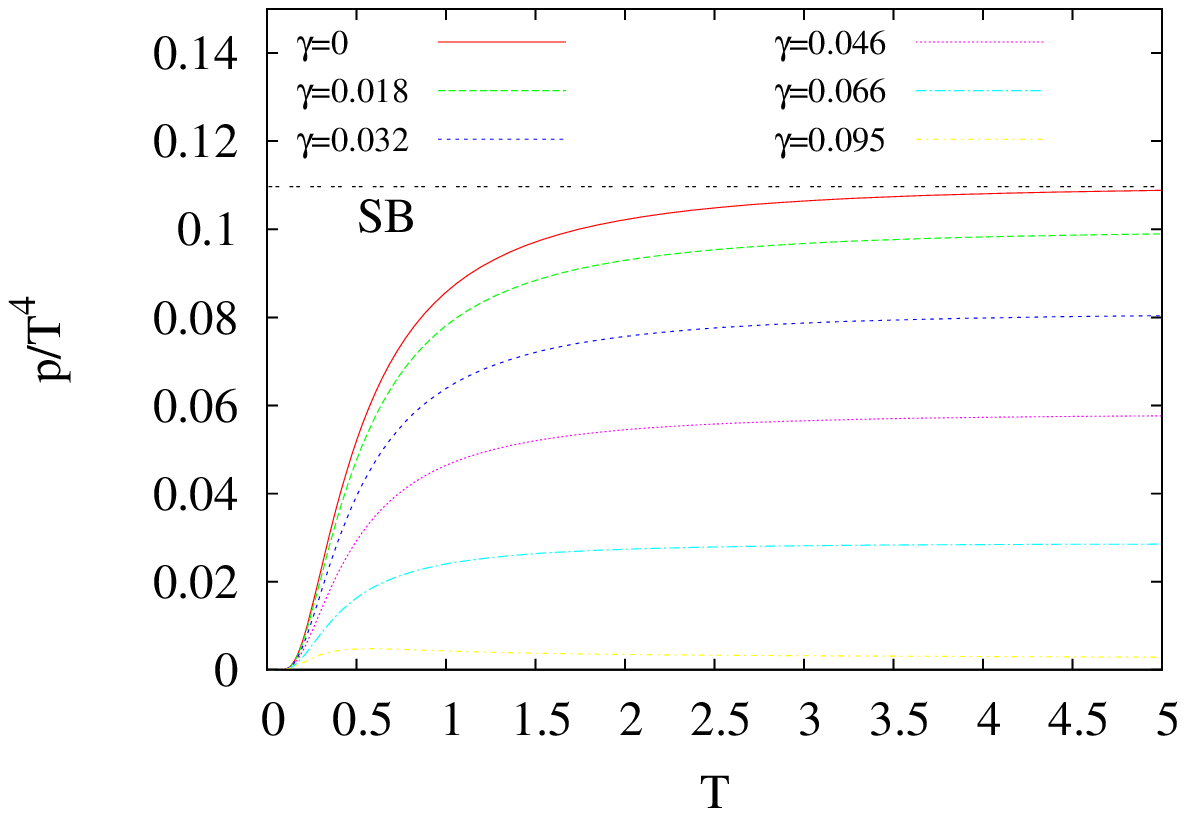}\\
  \hspace*{3em}a.)\hspace*{5cm}b.)
  \caption{a.) Spectral functions consisting of a single peak with
    different width parameters at $m=1$ and a 2-particle cut starting
    at threshold $m_{th}=2$. The curves are rescaled for better
    visibility. b.) The width dependence of the pressure.}
  \label{fig:Hmelt}
\end{figure}

First let us look at the result of the first scenario. For various
$\gamma$ values we can see the shape of the spectral
functions\footnote{Note that the curves are rescaled for better
  visibility} in Fig.~\ref{fig:Hmelt}/a.
We see that the peak has larger and larger width which results in
the suppression of the separation between the peaks and the
continuum. Still, the peak stays clearly identifiable.

After evaluating the integrals of \eqref{eq:master}, the corresponding
pressure data can be seen in Fig.~\ref{fig:Hmelt}/b. We see, how in
the course of increasing the overlap between the bound state and the
continuum the pressure decreases. We can observe that, despite the
fact that the peak cannot be missed in the spectrum, it is not a
thermodynamical degree of freedom beyond a certain width. Considering
the two widest peaks of the spectrum in Fig.~\ref{fig:Hmelt}/a, for
$\gamma=0.045$ we have half of the pressure yet, for $\gamma=0.137$
the pressure practically vanishes. The rule of thumb which can be
deduced from here is similar to the findings of \cite{Jakovac:2012tn}:
if $\gamma/\Delta m$ exceeds about 10\%, the peak cannot be considered
as a standalone, ideal thermodynamical degree of freedom, although
dynamically (eg. in linear response theory) one still can identify it.

\begin{figure}[htbp]
  \centering
  \includegraphics[height=5cm]{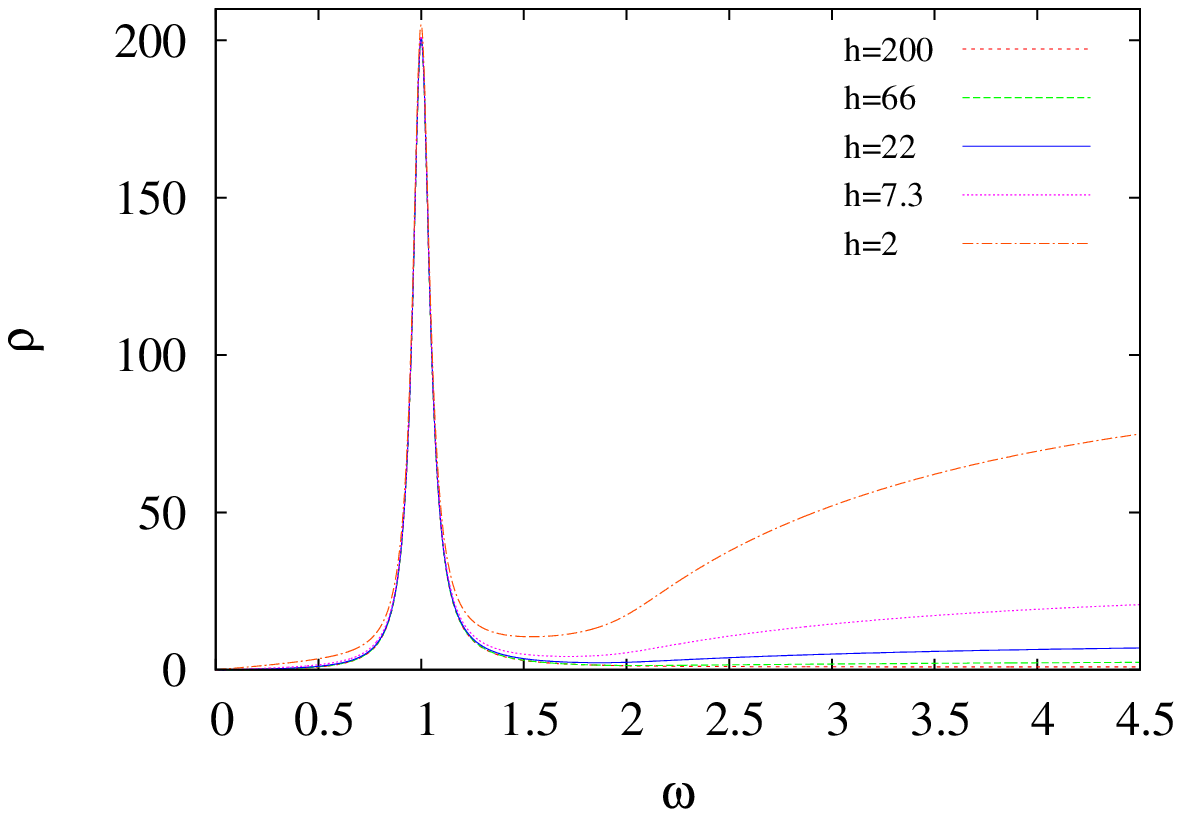}
  \includegraphics[height=5cm]{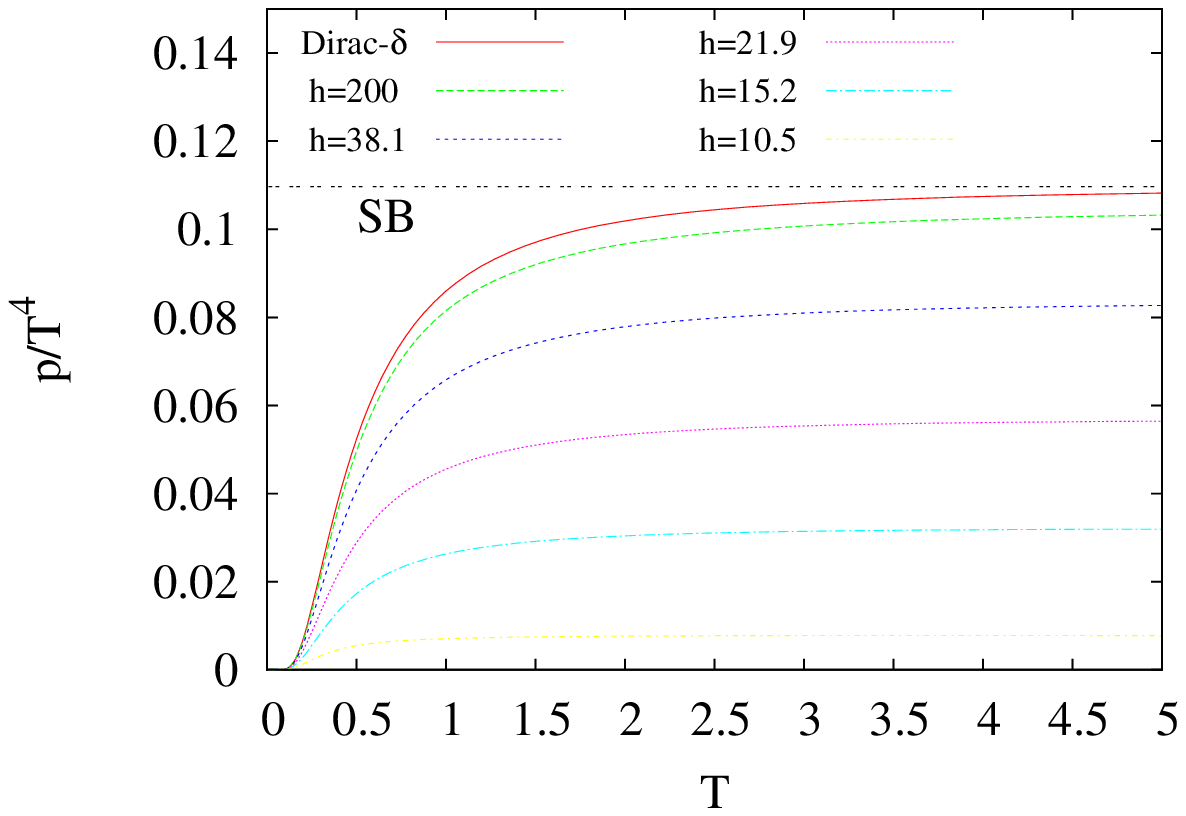}\\
  \hspace*{3em}a.)\hspace*{5cm}b.)
  \caption{a.) Spectral functions consisting of a single peak at $m=1$
    and $\gamma=0.05$, and a 2-particle cut with varying relative
    height starting at threshold $m_{th}=2$. $h$ is the peak height of
    \eqref{eq:Lorformula}. The curves are rescaled to keep the
    quasiparticle peak unchanged. b.) The dependence of the pressure
    on the continuum height.}
  \label{fig:Helim}
\end{figure}

In case of the second scenario we consider spectral functions plotted
in Fig.~\ref{fig:Helim}/a. Here all plots contain a quasiparticle with
the same width ($\gamma=0.05$ in the present case), but with
decreasing relative peak height (the plots are rescaled to have a
constant peak height with increasing value of the continuum). We see
that although the peak width is constant, the overlap is still getting
larger because of the increasing relative weight of the continuum part.

The pressure computed from these spectral functions, using
\eqref{eq:master}, can be seen in Fig.~\ref{fig:Helim}/b. The curves
are very similar to the previous scenario: increasing relative
continuum height leads to decreasing pressure. We can observe also
here the fact that, even when the peak is clearly identifiable in the
spectrum, and thus dynamically it still dominates the long time linear
response, in thermodynamics its contribution is already negligible.

We should emphasize that we have tried several parameter sets in the
aforementioned scenarios and also various shapes for the spectral
functions. The qualitative behavior, however, still remained the same
as in these cases.

\subsection{Effective number of thermodynamical degrees of freedom}

Our goal, of course, is to tell something about the thermodynamics of
QCD. To this end we want to draw some robust consequences of the
discussed toy model calculations which are independent of the detailed
mechanisms.

\begin{figure}[htbp]
  \centering
  \begin{minipage}[c]{6.5cm}
    \includegraphics[height=5cm]{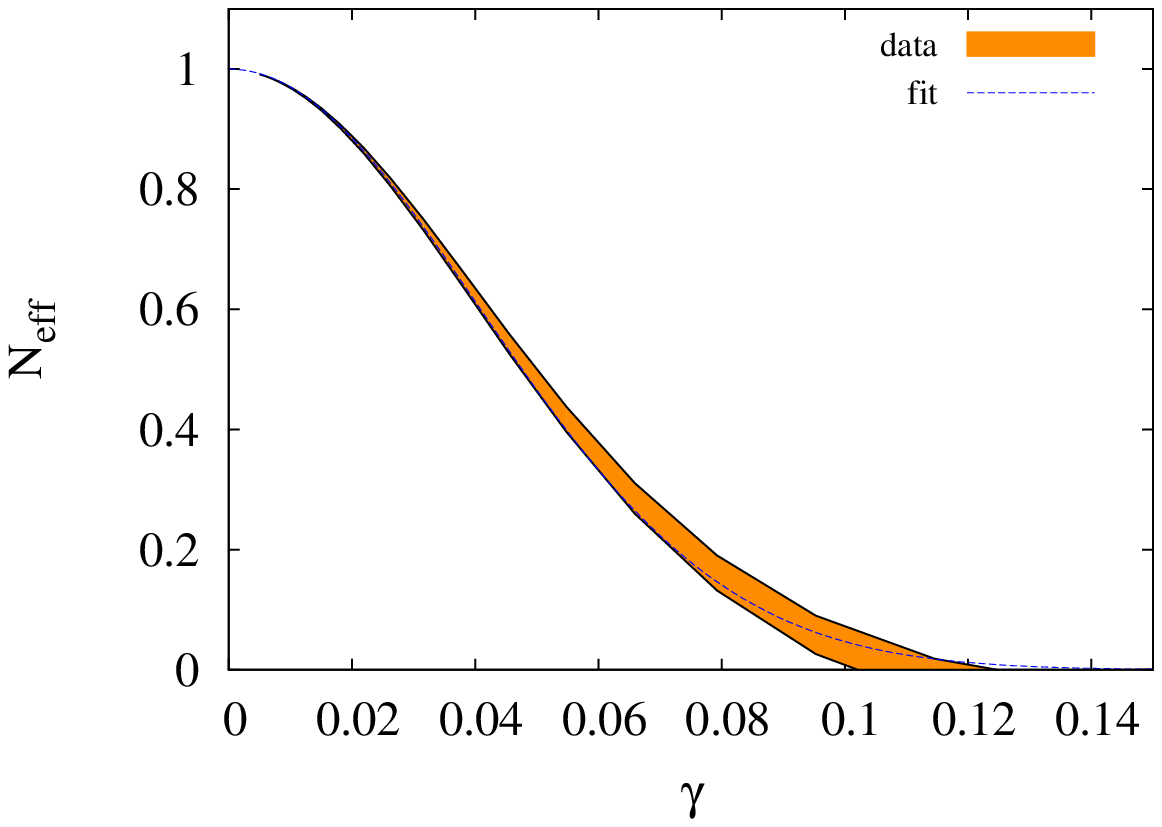}\\
    \hspace*{3em}a.)    
  \end{minipage}
  \hspace*{2em}
  \begin{minipage}[c]{6.5cm}
    \includegraphics[height=5.35cm]{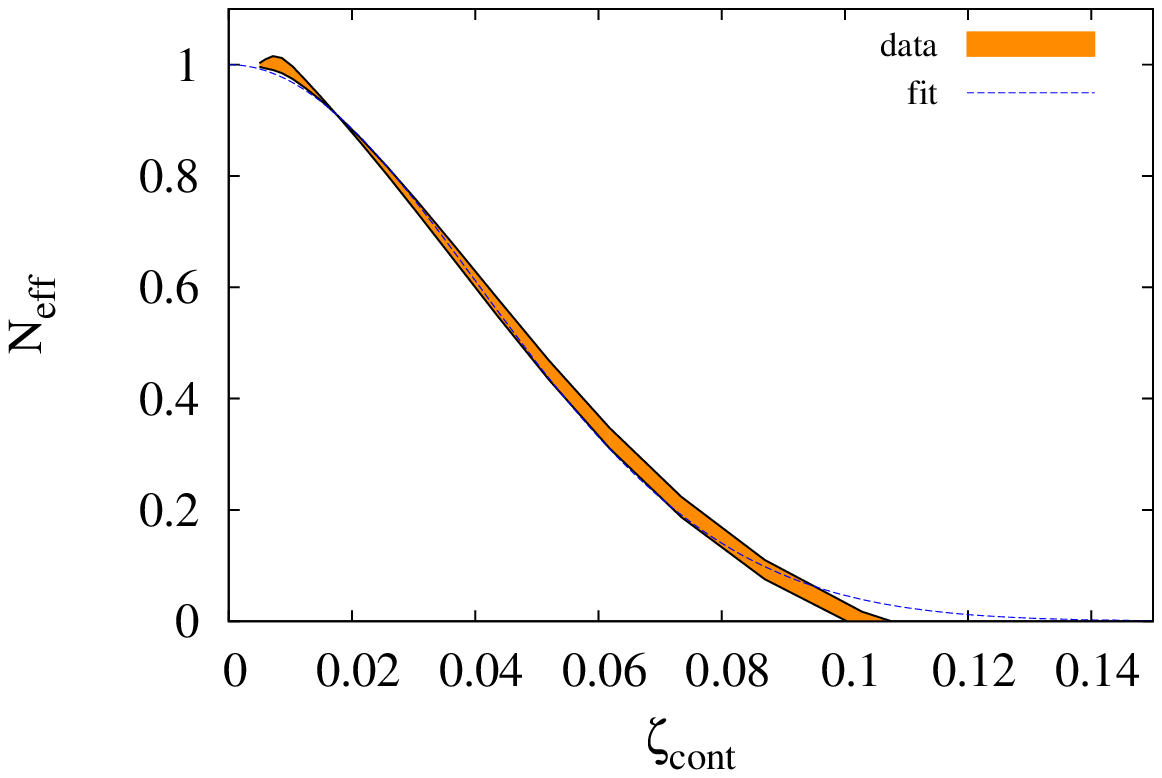}\\[-1.2em]
    \hspace*{3em}b.)    
  \end{minipage}
  \caption{Effective number of thermodynamical degrees of freedom in
    the studied temperature range, for a.) Scenario 1, b.) Scenario
    2. The shaded regions represent temperature dependence of the
    definition. The fit functions were Gaussian.}
  \label{fig:reduction}
\end{figure}

A common feature of all cases was the reduction of the number of
thermodynamical degrees of freedom. As we have seen, this quantity is
not closely related to the observability of the quasiparticle peaks in
the spectrum, thermodynamically the quasiparticle excitations become
much earlier negligible. Therefore, we introduce the effective number of
thermodynamical DoF $N_{eff}$ which depends on the temperature and the
quasiparticle parameters (width, height), and is defined simply as
\begin{equation}
  \label{eq:neff}
  N_{eff}(T) = \frac{p(T)}{p_{ideal}(T)},
\end{equation}
where in the ideal case we have only a Dirac-delta peak with mass
$m=1$. Evaluating this expression for our model-scenarios resulted in
plots of Fig.~\ref{fig:reduction}. The plots are restricted to the
physically sensible positive pressure parts.
The control parameter in case of Scenario 2 was
\begin{equation}
  \zeta_{cont} =\frac1{1+h},
\end{equation}
where $h$ is the peak height. $1-\zeta_{cont}$ can be physically
interpreted as a quantity which is proportional to the wave function
renormalization of the quasiparticle peak.

The definition of $N_{eff}$ is, in principle, temperature dependent,
so if we plot it against $\gamma$ or $\zeta_{cont}$ we have several
values corresponding to the actual temperature choice. The shaded
regions in Fig.~\ref{fig:reduction} show the temperature variation of
$N_{eff}$. As we see it is rather moderate, and it makes it possible
to speak about the number of thermodynamical degrees of freedom
independently of the actual temperature.

We can also try to write up an analytic approximate formula for these
numerically determined curves. Surprisingly, it appears that a simple
Gaussian curve
\begin{equation}
  N_{eff}(x) = e^{-x^2/2\sigma^2}
\end{equation}
works excellently in both cases, where $x$ is the corresponding
control parameter. In the actual calculations in the first scenario
the variance was $\sigma_\gamma=0.04$, in the second one
$\sigma_{\zeta_{cont}}=0.04$, too (a mere numerical coincidence).

\section{QCD thermodynamics}

Now let us try to draw the consequences of the above studies for
QCD. We emphasize it again that, according to the Hagedorn picture,
the thermal properties of QCD are determined by the statistical
ensemble of all hadrons. So we have to determine the thermal behavior
of an ``average'' hadron.

The fact that in both of our model-scenarios we found Gaussian
reduction factor for the effective number of dof, suggests that in the
QCD case we should also take Gaussian trial functions for this
quantity. There are two effects which we have to take into
account. The first is that the relative weight of the continuum, here
characterized by the pure number $\zeta_{cont}$, increases with the
number of the decay channels. For increasing masses, therefore, we
should have larger $\zeta_{cont}$ values. If $\zeta_{cont}$ is at
least linearly proportional to the hadron mass, then from the Gaussian
form we find a regularizing factor $\sim e^{-\# m^2}$ (or even faster
decreasing with $m$). This results that the most massive hadronic
states will never appear as separate full-fledged thermodynamical
degrees of freedom. In our calculation -- since the exact form of
$\zeta_{cont}$ is not known -- we take this effect into account by
introducing a maximal number of hadronic resonances. Similar technique
is proposed in \cite{Cleymans:2011fx}. We have checked that the final
results do not depend too much on the exact choice of this number.

The other effect is the temperature dependence of the width of the
bound states. Here we assume that the partial pressure \emph{of all
  hadrons} decreases in the same way. Physically we can motivate this
choice by the observation that the typical width of the hadrons at
zero temperature is in the same range of ${\cal
  O}(100)\,\mathrm{MeV}$, which makes it probable that the physics
which governs the width of the hadrons is similar, too. The fact that
the light and strange-quark susceptibilities are rescaled versions of
each other \cite{Bellwied:2013cta}, also supports the idea of a common
melting mechanism.

Then, using \eqref{eq:neff} and the fit of Fig.~\ref{fig:reduction} we
may consider a simple Ansatz
\begin{equation}
  \label{eq:meltansatz}
  p(T) = p_{Hag}(T) N_{eff}(\gamma(T)) = p_{Hag}(T) e^{-\kappa\gamma^2(T)},
\end{equation}
where $p_{Hag}$ is the pressure coming from the ideal Hagedorn gas
spectrum.

The (average) hadron width function $\gamma(T)$ is not really known,
but we can make the following considerations. First of all the width
is not zero at $T=0$ for most cases. On the other hand dimensional
analysis suggests that at high temperatures $\gamma(T)\sim T$. In many
model calculations these two regimes are cleanly separated: for small
temperatures the width is nearly constant or just slightly temperature
dependent, while at higher temperatures we get into a linearly rising
regime. One finds this behavior in the $\Phi^4$ model in two-loop 2PI
approximation \cite{Jakovac:2006gi}, in QCD using sum rules
\cite{Dominguez:2007ic}, in the holographic approach
\cite{Colangelo:2009ra}, and it is consistent also with pion gas 2PI
calculations \cite{Riek:2004kx}. To mimic this behavior we will take
the following simple form\footnote{Of course, other parametrizations
  are also possible, and we tried some of them different formula, too,
  but the final outcome remained the same.} for the suppression factor
\begin{equation}
  \kappa\gamma^2(T) = \kappa^2_0 + \delta\kappa^2(T),\qquad \delta\kappa(T)
  = \Theta(T-T_0) (T-T_0)/T_n.
\end{equation}
Here $\kappa_0$ represents the effect of the zero temperature
suppression, $T_0$ is the temperature where the linear regime starts,
$T_n$ is connected to the variance of the Gaussian.

\begin{figure}[htbp]
  \centering
  \includegraphics[height=5cm]{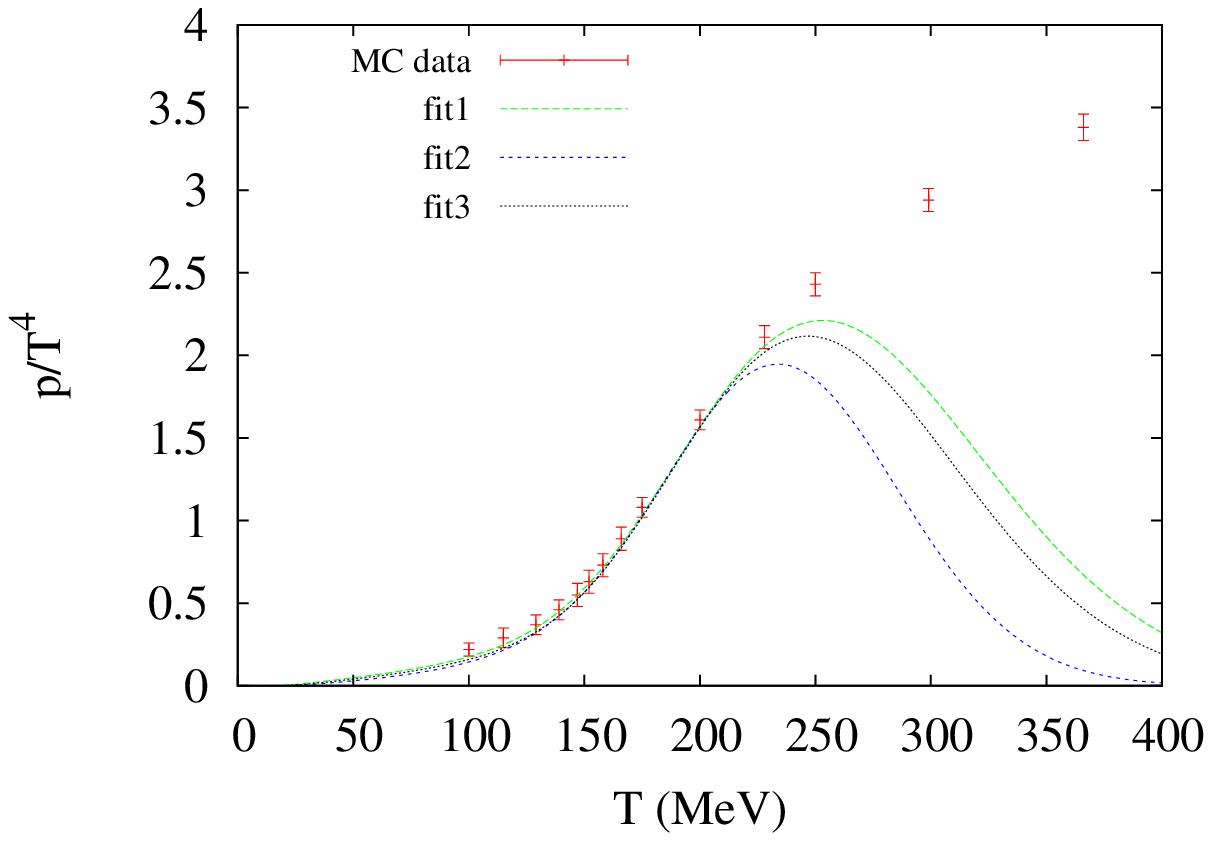}
  \includegraphics[height=5cm]{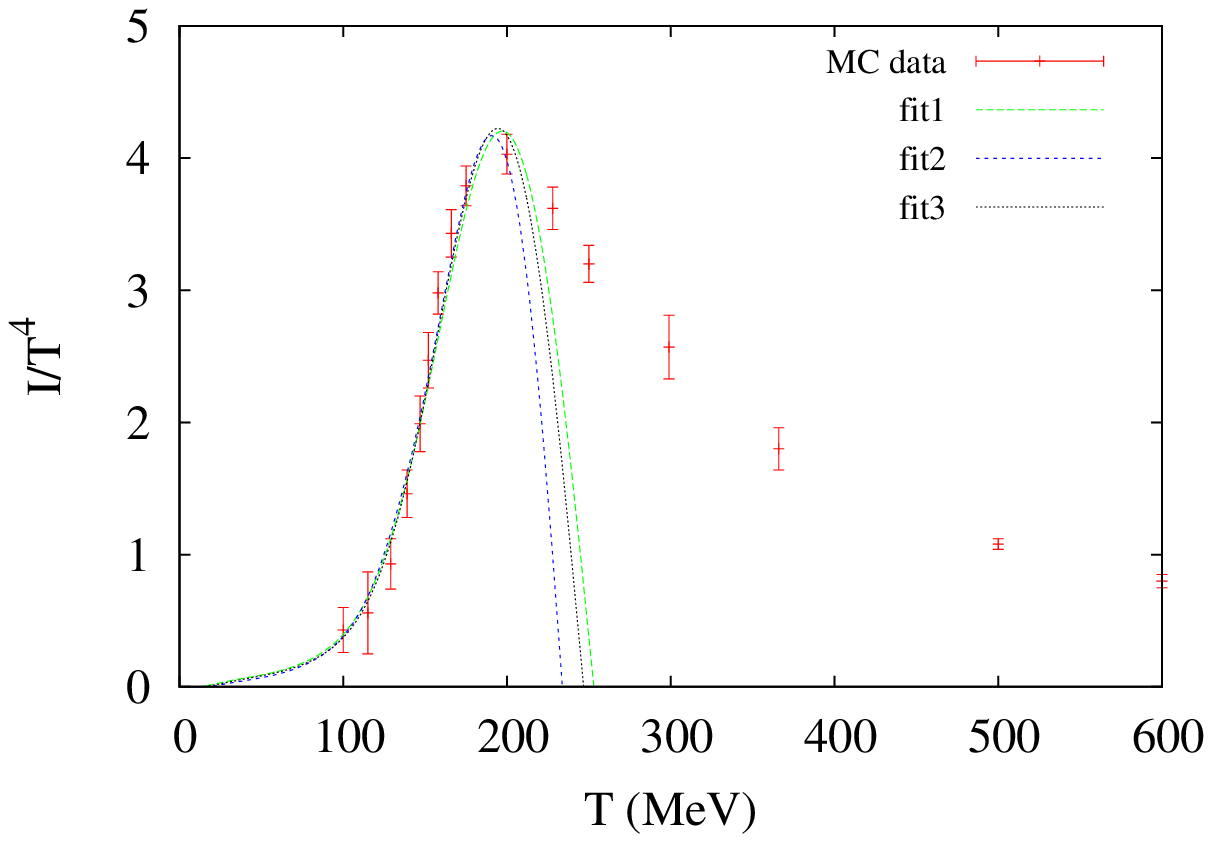}\\
  \hspace*{3em}a.)\hspace*{5cm}b.)
  \caption{Effect of the reduction of the number of degrees of freedom
    on the Hagedorn spectrum: a.) pressure, b.) interaction
    measure. The parameters of the fits is contained in
    \ref{tab:fits}.}
  \label{fig:fits}
\end{figure}

There are three parameters in the above formula which are determined
by fitting the MC trace anomaly data. Interestingly the fits are quite
robust in the sense that it is relatively easy to find reasonably
fitting solutions, with rather different values of the parameters. For
example, varying the Hagedorn temperature $T_H$, the starting mass $M$
and the zero temperature damping $\kappa_0^2$ all change the curve
similarly, and affect mainly the low temperature part of the
curve. $T_0$ plays a not too important role, and finally $T_n$ is
determined by the height of the trace anomaly curve. We plot some of
the results in Fig.~\ref{fig:fits}.
The fit parameters can be seen in Table \ref{tab:fits}.
\begin{table}[htbp]
  \centering
  \begin{tabular}[c]{||c|c|c|c|c|c|c||}
    \hline
    fit \# & $N$ & $T_H$ (MeV) & $M$ (MeV) & {\quad$\kappa_0^2$\quad}
    & $T_0$ (MeV) & $T_n$ (MeV) \cr
    \hline
    1      & $20000$ & $195$& $115$ & $0$  & $0$   & $158$ \cr
    2      & $20000$ & $220$& $140$ & $0.2$&$125$  & $\;\;93$ \cr
    3      & $40000$ & $190$& $130$ & $0$  & $0$   & $147$ \cr
    \hline
  \end{tabular}
  \caption{The fit parameters}
  \label{tab:fits}
\end{table}

The first fact that we should pin down is that indeed our effective
model is capable to provide a hadronic partial pressure which is smaller
than the complete pressure, as we should require.

If we look through the parameter choices of the different fits in
Table \ref{tab:fits}, we see that the first and third fits use a
minimal setup ($\kappa_0^2=0$ and $T_0=0$), the difference between
these two cases is the number of hadronic resonances we have taken
into account. Thus these fits are pure Gaussian $N_{eff}(T) =
e^{-T^2/T_n^2}$. Moreover, the lightest hadron masses were $M=115$ MeV
and $130$ MeV, which are consistent with a smeared out pion
mass. Still this very naive approach reproduces nicely the details of
both the measured hadronic pressure and the trace anomaly (interaction
measure)! The second row of Table~\ref{tab:fits} is a typical fit with
all parameters involved. We observe just very little difference
between the corresponding curves in Fig.~\ref{fig:fits}. We must also
emphasize that other thermodynamical quantities can be computed from
these two, eg. $\ep = I+3p$ and $Ts = I+4p$ are the energy density and
entropy density, respectively. Therefore, we can state that these
curves provide fits for the complete QCD thermodynamics in a certain
temperature range.

\begin{figure}[htbp]
  \centering
  \includegraphics[height=5cm]{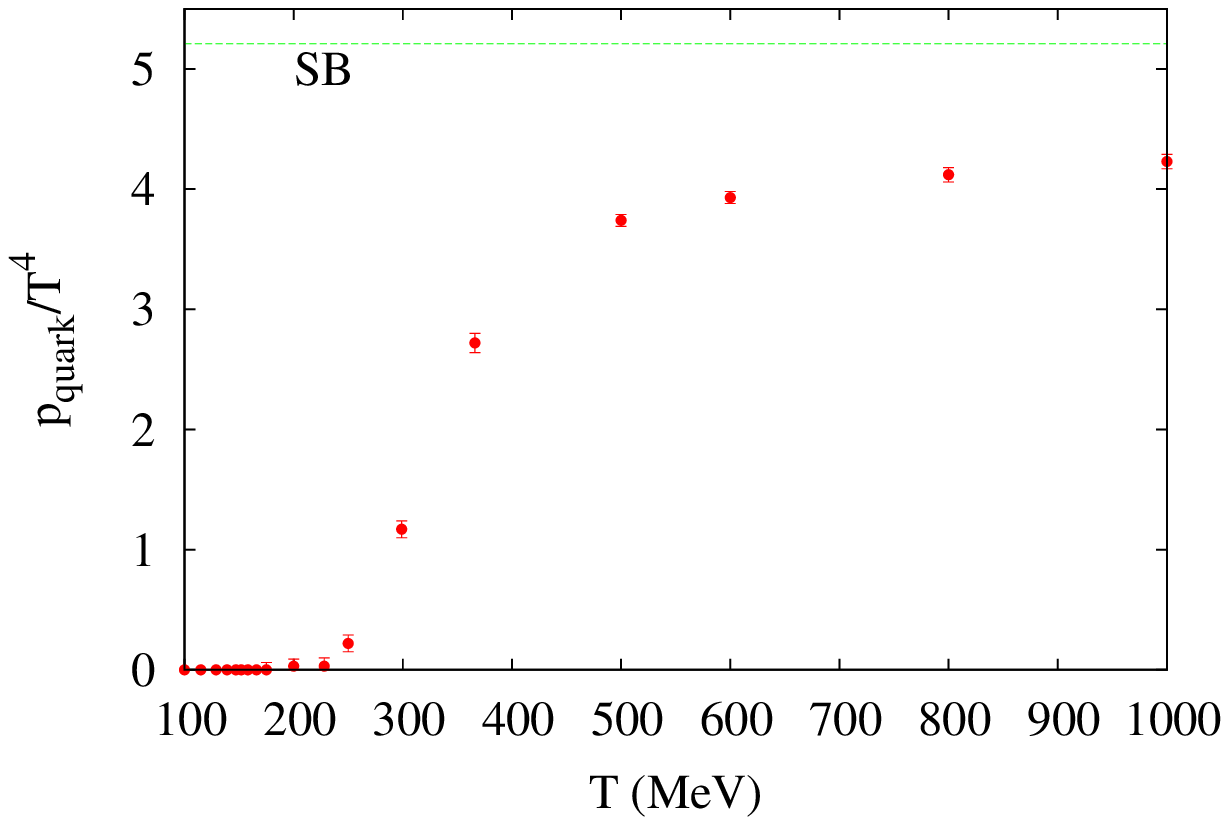}
  \includegraphics[height=5cm]{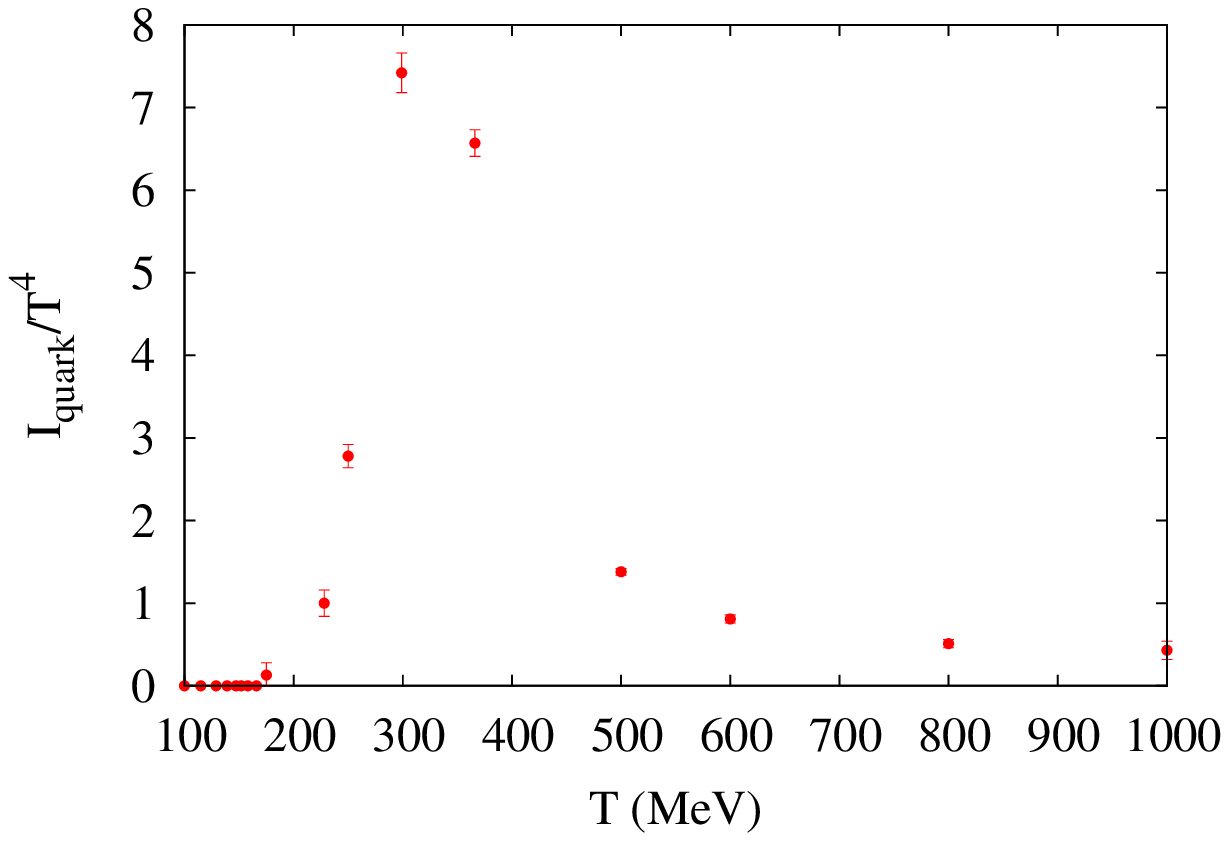}\\
  \hspace*{3em}a.)\hspace*{5cm}b.)
  \caption{The pressure and trace anomaly (interaction measure) of the
  QGP, according to our model predictions and MC data. SB means the
  asymptotic Stefan-Boltzmann limit.}
  \label{fig:QGPpI}
\end{figure}

The most striking property of these curves is that they can describe
full QCD pressure and interaction measure up to about $200-250$ MeV
(depending on the parametrization of the width). Since the full
pressure is the sum of the partial pressures of the hadronic and QGP
quantum channels $P_{tot} = P_{hadrons} + P_{QGP}$, we can read out
the QGP pressure and interaction measure coming from this model. We
plotted these quantities in Fig.~\ref{fig:QGPpI}, (using the first row
of Table~\ref{tab:fits}). As we can see, the pressure of the QGP
disappears below of about $250$ MeV. The physical reason is the
appearance of hadronic bound states which results in a drastic
decrease of the mean free path and lifetime of the QGP excitations.

\section{Conclusions}

In this paper we have studied the process of hadron melting based on
the method of Ref.~\cite{Jakovac:2012tn}. To describe real hadrons in
a thermal environment we have considered two model-scenarios for the
spectral function which shows the presence of resonances and a
continuum in a given quantum channel. The two cases differ in the way
of choosing the control parameter. We have found in all cases that
increasing the deviation from the single stable particle spectral
function resulted in a decreasing pressure, i.e. the quasiparticle,
represented by the peak ``melts''. The effective thermodynamical dof,
defined by the ratio of the actual pressure and the free particle
pressure, was determined numerically, and a Gaussian was fitted
to it to a very good precision.

These results then were adapted for QCD. We used a simple model, where
the masses of the hadronic resonances were distributed by Hagedorn
distribution, and hadron melting was taken into account by a Gaussian
width dependence, suggested by the model-scenarios. The width itself,
in the simplest case, was just approximated by $\gamma^2(T)=\gamma_0^2
+ c T^2$. As a result we could fit the MC results for QCD
thermodynamics (pressure and trace anomaly) up to $T'_c\approx 200$ -
$250$ MeV with few parameters. We also determined the QGP pressure
which, by lowering the temperature, vanishes in this temperature
range.

What do these results tell us? We must say that, according to the
statistical hadron picture combined with the spectral analysis of the
hadron resonances, \emph{the hadron - QGP transition occurs in the
  temperature range of} $T'_c\sim 200$-$250$ MeV. The reported
crossover at the lower temperature $T_c=156$ MeV characterizes the
hadron melting process: it is the starting point of rapid hadron
melting. It can also be considered as a limiting temperature
separating the regimes where the naive, free HRG model works from the
one where the hadron width (more precisely the complete spectral
function) becomes an important factor, and no uncorrelated hadrons are
present any more. In this light the findings of
\cite{Bellwied:2013cta} that the strange and non-strange sectors
present different $T_c$, support the above picture, since the more
massive hadrons may melt later. But for a good while the effective
hadronic number of dof are still significantly higher than the number
of dof of QGP. The hadronic pressure shrinks only at about
$T'_c\approx 200-250$ MeV temperature to that small values, that the
QGP pressure becomes dominant.

Numerically it would be really challenging to find whether there is
any signal of any phase transition in the temperature regime $T'_c\sim
200$-$250$ MeV. It is presently a much less studied temperature range,
and from the existing data it is impossible to tell anything about it.

According to this proposal, the temperature range of $T\sim[156,250]$
should be treated by a correlated hadron gas model. In this model
perturbative calculations are possible (provided one uses the full
spectral function), while PQCD calculations are relevant from $T\sim
200$-$250$ MeV. To test this picture, one should calculate other
quantities, like thermodynamical observables at finite chemical
potential, or transport coefficients. These findings could be
relevant not just for collider physics, but, as in case of the
chemical potential dependence, also for astrophysics.

\section*{Acknowledgments}

The author thanks long and instructive discussions with
A. Patk\'os. He also acknowledges discussions with and important
remarks from T.S. B\'{\i}r\'o, Sz. Bors\'anyi, S.D. Katz, P. Petreczky
and Zs. Sz\'ep. This work is supported by the Hungarian Research Fund
(OTKA) under contract No. K104292.


\begin{thebibliography}{99}

\bibitem{HRG0} for a review see: P. Braun-Munzinger,
  K. Redlich, and J. Stachel, In Hwa, R.C. (ed.) et al.: \emph{Quark
    gluon plasma} 491-599, [nucl-th/0304013].


\bibitem{Andronic:2003} A. Andronic, P. Braun-Munzinger,
  K. Redlich and J. Stachel, Phys. Lett. B571, 36(2003)
  \eprint{nucl-th/0303036}

\bibitem{Karschetal} F. Karsch, K. Redlich and A. Tawfik,
  Eur. Phys. J. C29, 549(2003) \eprint{hep-ph/0303108}

\bibitem{Huovinen:2009yb}
  P.~Huovinen and P.~Petreczky,
  Nucl.\ Phys.\  A {\bf 837}, 26 (2010)
  [arXiv:0912.2541 [hep-ph]].


\bibitem{Borsanyi:2010cj} 
  S.~Borsanyi, G.~Endrodi, Z.~Fodor, A.~Jakovac, S.~D.~Katz, S.~Krieg, C.~Ratti and K.~K.~Szabo,
  JHEP {\bf 1011}, 077 (2010)
  [arXiv:1007.2580 [hep-lat]].

\bibitem{Bazavov:2013dta} 
  A.~Bazavov, H.~-T.~Ding, P.~Hegde, O.~Kaczmarek, F.~Karsch, E.~Laermann, Y.~Maezawa and S.~Mukherjee {\it et al.},
  arXiv:1304.7220 [hep-lat].

\bibitem{PDG} J. Beringer et al. (Particle Data Group),
  Phys. Rev. D86, 010001 (2012)

\bibitem{Aoki:2006we} 
  Y.~Aoki, G.~Endrodi, Z.~Fodor, S.~D.~Katz and K.~K.~Szabo,
  Nature {\bf 443}, 675 (2006)
  [hep-lat/0611014].

\bibitem{Hagedorn:1965st} 
  R.~Hagedorn,
  Nuovo Cim.\ Suppl.\  {\bf 3}, 147 (1965);

\bibitem{Datta:2003ww} 
  S.~Datta, F.~Karsch, P.~Petreczky and I.~Wetzorke,
  Phys.\ Rev.\ D {\bf 69}, 094507 (2004)
  [hep-lat/0312037].

\bibitem{Umeda:2002vr} 
  T.~Umeda, K.~Nomura and H.~Matsufuru,
  Eur.\ Phys.\ J.\ C {\bf 39S1}, 9 (2005)
  [hep-lat/0211003].

\bibitem{Asakawa:2003re} 
  M.~Asakawa and T.~Hatsuda,
  Phys.\ Rev.\ Lett.\  {\bf 92}, 012001 (2004)
  [hep-lat/0308034].

\bibitem{Jakovac:2006sf} 
  A.~Jakovac, P.~Petreczky, K.~Petrov and A.~Velytsky,
  Phys.\ Rev.\ D {\bf 75}, 014506 (2007)
  [hep-lat/0611017].

\bibitem{Petreczky:2012ct} 
  P.~Petreczky,
  J.\ Phys.\ Conf.\ Ser.\  {\bf 402}, 012036 (2012)
  [arXiv:1204.4414 [hep-lat]].

\bibitem{Bellwied:2013cta} 
  R.~Bellwied, S.~Borsanyi, Z.~Fodor, S.~D.~Katz and C.~Ratti,
  arXiv:1305.6297 [hep-lat].

\bibitem{Heinz:2013th} 
  U.~W.~Heinz and R.~Snellings,
  arXiv:1301.2826 [nucl-th].

\bibitem{Blaschke:2003ut} 
  D.~B.~Blaschke and K.~A.~Bugaev,
  Fizika B {\bf 13}, 491 (2004)
  [nucl-th/0311021].

\bibitem{Biro:2006sv} 
  T.~S.~Biro and J.~Zimanyi,
  Phys.\ Lett.\ B {\bf 650}, 193 (2007)
  [hep-ph/0607079].

\bibitem{Biro:2006zy} 
  T.~S.~Biro, P.~Levai, P.~Van and J.~Zimanyi,
  J.\ Phys.\ G {\bf 32}, S205 (2006)
  [hep-ph/0605274].


\bibitem{Ivanov:1998nv} 
  Y.~.B.~Ivanov, J.~Knoll and D.~N.~Voskresensky,
  Nucl.\ Phys.\ A {\bf 657}, 413 (1999)
  [hep-ph/9807351].

\bibitem{Ivanov:1999tj} 
  Y.~.B.~Ivanov, J.~Knoll and D.~N.~Voskresensky,
  Nucl.\ Phys.\ A {\bf 672}, 313 (2000)
  [nucl-th/9905028].

\bibitem{Peshier:2005pp} 
  A.~Peshier and W.~Cassing,
  Phys.\ Rev.\ Lett.\  {\bf 94}, 172301 (2005)
  [hep-ph/0502138].

\bibitem{RD} R.F. Dashen and R. Rajaraman, Phys.Rev. \textbf{D10}
  (1974) 694; Phys.Rev. \textbf{D10} (1974) 708

\bibitem{Scon} H. Feshbach, Ann. Phys. 43, 110 (1967),
  L. Rosenfeld, Acta Phys. Polonica A38, 603 (1970).

\bibitem{Svec} 
  M.~Svec,
  Phys.\ Rev.\ D {\bf 64}, 096003 (2001)
  [hep-ph/0009275].

\bibitem{Jakovac:2012tn} 
  A.~Jakovac,
  Phys.\ Rev.\ D {\bf 86}, 085007 (2012)
  [arXiv:1206.0865 [hep-ph]].

\bibitem{Broniowski:2004yh} 
  W.~Broniowski, W.~Florkowski and L.~Y.~.Glozman,
  Phys.\ Rev.\ D {\bf 70}, 117503 (2004)
  [hep-ph/0407290].

\bibitem{NoronhaHostler:2012ug} 
  J.~Noronha-Hostler, J.~Noronha and C.~Greiner,
  Phys.\ Rev.\ C {\bf 86}, 024913 (2012)
  [arXiv:1206.5138 [nucl-th]].

\bibitem{Cleymans:2011fx} 
  J.~Cleymans and D.~Worku,
  Mod.\ Phys.\ Lett.\ A {\bf 26}, 1197 (2011)
  [arXiv:1103.1463 [hep-ph]].

\bibitem{Jakovac:2006gi} 
  A.~Jakovac,
  Phys.\ Rev.\ D {\bf 76}, 125004 (2007)
  [hep-ph/0612268].

\bibitem{Dominguez:2007ic} 
  C.~A.~Dominguez, M.~Loewe and J.~C.~Rojas,
  JHEP {\bf 0708}, 040 (2007)
  [arXiv:0707.2844 [hep-ph]].

\bibitem{Colangelo:2009ra} 
  P.~Colangelo, F.~Giannuzzi and S.~Nicotri,
  Phys.\ Rev.\ D {\bf 80}, 094019 (2009)
  [arXiv:0909.1534 [hep-ph]].

\bibitem{Riek:2004kx} 
  F.~Riek and J.~Knoll,
  Nucl.\ Phys.\ A {\bf 740}, 287 (2004)
  [nucl-th/0402090].

\end{thebibliography}
\end{document}